\documentclass[reprint,amsmath,amssymb,aps,pra,reprint,
 amsmath,amssymb,
 pra,superscriptaddress]{revtex4-2}

\usepackage{mypackages}
\usepackage{tikz}
\usetikzlibrary{shapes.geometric, arrows, positioning}

\tikzstyle{process} = [rectangle, rounded corners, minimum width=3.5cm, minimum height=1.2cm, text centered, draw=black, align=center,font=\Huge]
\tikzstyle{arrow} = [thick,->,>=stealth, line width=0.8mm, scale=3]

\usepackage{amscd}
\usepackage{textcomp}
\usepackage[thinlines]{easybmat}
\usepackage{multirow,bigdelim}
\usepackage{adjustbox} 

\begin{document}

\title{Automorphism Ensemble Decoding of Quantum LDPC Codes}

\author{Stergios Koutsioumpas}
\thanks{\ These authors contributed equally to this work. Corresponding author: \url{s.koutsioumpas@ucl.ac.uk}}
\affiliation{Department of Physics \& Astronomy, University College London, London, WC1E 6BT, United Kingdom}
\author{Hasan Sayginel}
\thanks{\ These authors contributed equally to this work. Corresponding author: \url{s.koutsioumpas@ucl.ac.uk}}
\affiliation{Department of Physics \& Astronomy, University College London, London, WC1E 6BT, United Kingdom}
\affiliation{National Physical Laboratory, Teddington, TW11 0LW, United Kingdom}

\author{Mark Webster}
\affiliation{Department of Physics \& Astronomy, University College London, London, WC1E 6BT, United Kingdom}

\author{Dan E Browne}
\affiliation{Department of Physics \& Astronomy, University College London, London, WC1E 6BT, United Kingdom}

\date{\today}

\begin{abstract}
We introduce AutDEC, a fast and accurate decoder for quantum error-correcting codes with large automorphism groups. 
Our decoder employs a set of automorphisms of the quantum code and an ensemble of belief propagation (BP) decoders.
Each  BP decoder is given a syndrome which is transformed by one of the automorphisms, and is run in parallel.
For quantum codes, the accuracy of BP decoders is limited because short cycles occur in the Tanner graph and our approach mitigates this effect.
We demonstrate decoding accuracy comparable to BP-OSD-0 with a lower time overhead for Quantum Reed-Muller (QRM) codes in the code capacity setting, and Bivariate Bicycle (BB) codes under circuit level noise. 
We provide a Python repository for use by the community and the results of our simulations. 
\end{abstract}

\maketitle


\section{\label{sec:introduction}Introduction}
Quantum low-density parity-check (LDPC) codes have the potential to significantly reduce the resource overhead required for fault-tolerant quantum computation, offering compelling advantages such as high encoding rates, robust distance scaling, and favourable error thresholds \cite{breuckmann_quantum_2021}. 
Whilst belief propagation (BP) decoders are efficient for use with classical LDPC codes, the presence of short cycles in the Tanner graphs of quantum LDPC codes mean that BP decoders are not always effective for quantum codes. 

Several approaches have been suggested to facilitate the application of BP decodes to quantum LDPC codes such as Ordered Statistics Decoding (OSD) \cite{panteleev_degenerate_2021}, Symmetry Breaking (SymBreak) \cite{yin_symbreak_2024},  Ambiguity Clustering \cite{wolanski_ambiguity_2025}, and Localized Statistics Decoding (LSD) \cite{hillmann_localized_2024}. 
Whilst these approaches improve the accuracy of BP decoding, they involve increased time overhead. 
For practical applications, the performance of the decoder must match the speed and accuracy requirements of the quantum hardware to avoid backlog problems \cite{terhal_backlog}. 

In this work we introduce the AutDEC \cite{Sayginel_AutDEC_2025} Automorphism Ensemble decoder for quantum codes. A code automorphism is a permutation of the bits of a code that preserves its span. 
Our decoder employs a set of automorphisms of the quantum code and an ensemble of BP decoders.
Each  BP decoder is given a syndrome which is transformed by one of the automorphisms.
In general, the corrections calculated for each transformed syndrome are distinct and this usually bypasses issues with short cycles.
We choose the highest probability correction from the ensemble for the final correction,  improving the accuracy compared to simple BP decoding.
Crucially, the BP decoders are run in parallel and so the time overhead is the same as running a single BP decoder.

Whilst automorphism ensemble decoders have been used for classical codes, this is the first time they have been used for quantum error-correcting codes.
We demonstrate decoding accuracy comparable to BP-OSD-0 with a lower time overhead for Quantum Reed-Muller (QRM) codes in the code capacity setting, and Bivariate Bicycle (BB) codes under circuit level noise. 

We present simulation results that showcase the decoder's performance and provide an open-source implementation of our methods using the open-source  Bliss \cite{junttila_engineering_2007, junttila_conflict_2011} and igraph \cite{Csardi_The_igraph_software_2006} packages to calculate automorphisms.

The structure of the paper is as follows. In Section \ref{sec:motivation} we provide background on the motivation for the algorithm. 
In Section \ref{theory} we give background on stabiliser error-correcting codes, belief propagation decoding, code and check matrix automorphisms. We introduce the automorphism ensemble decoder \enquote{AutDEC} in \ref{sec:autdec}, its rationale and a detailed description of the method. 
Finally,  in \ref{sec:results} we provide simulation results for Quantum Reed Muller Codes on the code capacity depolarising noise model and Bivariate Bicycle Codes under circuit-level noise.

\section{Motivation}\label{sec:motivation}

Classical LDPC decoding has seen remarkable progress, with algorithms such as belief propagation (BP) showing to be highly effective in approximating maximum-likelihood (ML) solutions with low computational cost \cite{pearl_reverend_1982, mackay_near_1996, mceliece_turbo_1998}.  
Translating these successes directly to the quantum domain with QLDPC codes presents significant new challenges.
The presence of  short cycles in the Tanner graph of quantum codes causes problems for message passing algorithms such as BP \cite{panteleev_degenerate_2021}. 
This complicates the decoding task and often necessitates post-processing beyond BP \cite{roffe_decoding_2020}. 
Techniques like Ordered Statistics Decoding (OSD) \cite{panteleev_degenerate_2021}, Symmetry Breaking (SymBreak) \cite{yin_symbreak_2024}, Ambiguity Clustering \cite{wolanski_ambiguity_2025}, and Localized Statistics Decoding (LSD) \cite{hillmann_localized_2024} have emerged to solve the problem, but come at the cost of increased time complexity compared to standard BP.

Despite these algorithmic innovations, achieving the necessary sustained, high-throughput decoding for practical quantum LDPC-based quantum computation remains a significant challenge.  Recent experimental demonstrations \cite{zhao_realization_2022, acharya_quantum_2024, lacroix_scaling_2024, caune_demonstrating_2024,rodriguez_experimental_2024, berthusen_experiments_2024} have highlighted the difficulty of the so-called \enquote{throughput problem} in error correction. As quantum systems become larger and operate at faster cycle times, they generate error syndromes at increasing rates.  The decoder must be capable of processing these syndromes and delivering corrections sufficiently quickly to keep pace with the ongoing quantum computation.  If the decoding process is too slow, or if the decoder's processing rate is insufficient to handle the incoming syndrome data stream, a computational backlog inevitably arises \cite{caune2024demonstratingrealtimelowlatencyquantum}.  


Automorphism ensemble decoding for classical Reed-Muller codes was introduced by Geiselhart et al in \cite{geiselhart_automorphism_2021}. 
This work demonstrated that leveraging the analytically known automorphism group of these codes \cite{abbe_reedmuller_2021} leads to effective decoding using BP. 
Automorphism ensemble decoding has been found effective when applied to classical LDPC codes \cite{geiselhart_ldpc_automorphism,krieg_comparative_2024,krieg_comparative_2024}. 

In the quantum setting, the application of ensemble decoding techniques is emerging as a potential candidate in different forms. 
Initial investigations have considered surface codes, often in conjunction with matching decoders \cite{shutty_efficient_2024, jones_improved_2024}. 
Related investigations include joint code-decoder design for Multiple Bases Belief Propagation \cite{miao_joint_2024} and methods combining different decoders for enhanced performance \cite{sheth_neural_2020}.

\section{Background}\label{theory}
In this section we will briefly outline the theory of quantum stabiliser error-correcting codes and the important subclass of CSS codes. 
We then introduce belief propagation decoding which we use for each of the decoders in the ensemble approach.
Each decoder is given a syndrome to decode which has been deformed by applying a code or check matrix automorphism - we introduce these concepts in the final part of this section.

\subsection{Stabiliser Codes}
Stabiliser codes are a class of quantum error-correcting codes defined by a set of commuting, multi-qubit Pauli operators called the \textbf{stabiliser group}.  
This group, denoted $S$, is a subgroup of the Pauli group on $n$ physical qubits. 
The \textbf{code space} is the +1 eigenspace of all the stabiliser operators. 
If the stabiliser group $S$ is generated by a set of $r$ independent \textbf{stabiliser generators}, then the code has $k = n-r$ logical qubits. 
Errors are detected and corrected by measuring the stabiliser generators. 
The measurement outcomes, known as the \textbf{syndrome vector} $s$, are passed to a decoder which aims to calculate a Pauli operator $c$ which corrects back into the codespace.
The distance $d$ of a quantum error-correcting code is defined as the lowest weight of any Pauli operator that commutes with the stabiliser generators but is not in the stabiliser group.



In this paper, we consider error correction for the Calderbank, Shor, Steane or CSS codes \cite{calderbank_good_1996,steane_multiple-particle_1997}.
These are specified using two classical codes which represent stabiliser generators of either $X$-type or $Z$-type.
An  $[n,k_1,d_1]$ classical code $C_1$ with generator matrix $H_X$ and an $[n,k_2,d_2]$ classical code $C_2$ with generator matrix $H_Z$ satisfying $H_X H_{Z}^{T}=0$ define a  $[[n,k_1+k_2-n,d]]$  quantum code 
with $d=\text{min}(d_1,d_2)$.
To form the stabiliser generators, we consider row $i$ of $H_X$ a stabiliser generator of $X$-type which applies a Pauli $X$ operator on qubit $j$ if $H_X[i,j] =1$.
The stabiliser generators of $Z$-type are defined similarly.
For a more in-depth review of quantum code automorphisms in the stabiliser formalism, we refer the reader to \cite{calderbank1997, sayginel_fault-tolerant_2024}.

\subsection{Belief Propagation Decoder}\label{sec:decoding}
In this section, we outline the BP decoders which we use as components of our ensemble decoder.
Belief propagation (BP) is a heuristic decoding algorithm developed by Pearl in \cite{pearl_reverend_1982}. 
BP iteratively passes the received syndrome data between variable and check nodes of the Tanner graph, until a correction satisfying the syndrome is found or a pre-determined maximum number of iterations is reached.
BP is efficient for classical LDPC codes because they have sparse Tanner graphs. 
Even for large classical LDPC codes, BP is close to optimal and has almost linear runtime on the code's length \cite{kschischang_factor_2006}. 

Different message passing schedules exist for BP, such as the flooding or parallel schedule and the layered or serial schedule \cite{goldberger_serial_2008}. This affects the ordering of the updates of the nodes on the Tanner graph and can have an effect on the number of iterations required or the convergence in different cases. For our results, we used the parallel BP schedule from \cite{Roffe_LDPC_Python_tools_2022} in order to minimize latency.

\subsection{Automorphisms of Codes and Check Matrices}\label{sec:autsofcodes}
The AutDEC decoder uses a set of automorphisms of the quantum code to transform the syndrome for correction.
For CSS  codes, automorphisms are qubit permutations $A$ that preserve the stabiliser group:
\begin{align*}
    SA = S.
\end{align*}
An automorphism may send a stabiliser generator to a product of stabiliser generators, but the transformed generators will still generate the same stabiliser group.
In our methods, we consider the automorphisms of the $X$-checks $H_X$ and the $Z$-checks $H_Z$ of a CSS code separately.
Given a permutation automorphism $A$, we show in Algorithm \ref{alg:stab_map} how to find an $r\times r$ binary invertible matrix $U_A$ such that:
\begin{align}
    U_A H_X = H_X A.\label{UA}
\end{align}
Let $s$ be the syndrome vector which results from measuring the stabiliser generators corresponding to $H_X$.
The syndrome vector we would find if measuring the transformed stabiliser generators $H_XA$ is given by $U_A\cdot s$.
This transformed syndrome vector will be used as the input to our ensemble of decoders.

For small codes ($n<360$), we can use \textbf{code automorphisms} of the underlying classical codes $C_1$ and $C_2$ (generated by $H_X$ and $H_Z$ respectively). These can be found by Leon's algorithm  \cite{leon_computing_1982} using the method outlined in \cite{sayginel_fault-tolerant_2024}.
For larger codes, we instead consider automorphisms of the check matrices $H_X$ and $H_Z$.
Check matrix automorphisms are a subgroup of the code automorphism group.
This is because a check matrix automorphism maps each stabiliser generator to another, rather than to products of stabiliser generators.

We consider the automorphisms of $H_X$ and $H_Z$ separately, and these are calculated by mapping the check matrices to their Tanner graphs.
The \textbf{Tanner graph} of a classical code is a bipartite graph consisting of two sets of nodes: 
\begin{enumerate}
    \item the variable nodes $V=\{ v_i, i \in [1,...,n]\}$ which correspond to the columns or bits the code and
    \item the check nodes $C=\{c_i, i\in [1,...,n-k]\}$ which represent the rows or stabilisers (parity check equations) of the code. 
\end{enumerate}
There is an edge between nodes $v_j$ and $c_i$ if the $i$-th row of the check matrix has a $1$ in the $j$-th column. 

A \textbf{graph automorphism} $A$ of a graph $\mathcal{G}$ is a permutation of the vertices of $\mathcal{G}$ which preserves the connectivity of the graph.
That is, $(v_j,c_i)$ is an edge of $\mathcal{G}$ if and only if $(v_jA,U_Ac_i)$ for some permutation $U_A$ of the rows of the check matrix.
Finding automorphisms of a Tanner graph is very fast in practice even for large codes, and there are publicly-accessible packages for this including Nauty \cite{mckay_practical_2014} or Bliss \cite{junttila_engineering_2007, junttila_conflict_2011}.
 

\begin{example}
    The automorphism group of the $[[15,1,3]]$ code is of order $20160$ and is isomorphic to General Linear group of degree $4$ over the field of $2$ elements $GL(4,2)$ \cite{macwilliams_theory_1977}. The automorphism group of the $H_X$ and $H_Z$ check matrices of the code, however is of order $24$, and is isomorphic to the Symmetric group of $4$ elements $S_4$. 
\end{example}


\section{Automorphism Ensemble Decoder}\label{sec:autdec}
The AutDEC decoder employs a set of automorphisms of the quantum code and an ensemble of BP decoders.
Each  BP decoder is given a syndrome which is transformed by one of the automorphisms.
In general, the corrections calculated for each transformed syndrome are distinct and this usually bypasses issues with short cycles.
We choose the highest probability correction from the ensemble for the final correction,  improving the accuracy compared to simple BP decoding.
Crucially, the BP decoders are run in parallel and so the time overhead is the same as running a single BP decoder. A diagram of the method can be seen in Figure \ref{fig:AED-overview}.

\subsection{Rationale for AutDEC Decoder}
The effectiveness of our decoder relies on each BP decoder giving a different correction based on syndromes which have been transformed by code or check matrix automorphisms.
By deforming the syndrome and permuting the nodes of the Tanner graph for the code, the BP decoder is less likely to be adversely affected by short cycles in the Tanner graph. 
We illustrate this by showing how the decoder acts on the 15-qubit quantum Reed-Muller code.

\begin{example}\label{rm15BP}
   Consider a $Z$-error on the $15$-th qubit of the $[[15,1,3]]$ QRM code. As all $4$ $X$-checks have support on this qubit, a $4$ cycle is created, and BP will output the all $1$s correction, ie $Z^{\otimes 15}$. 

   Using the code automorphism $A=(2,9)(3,8)(4,15)(5,14)$, exchanges the $4$th and $15$th qubits, and thus only one check is satisfied. This yields the right correction, on the permuted basis, of $Z_{15}$.

   In Figure \ref{fig:bp_rm_15_aut} we show the action of the automorphism $A$ on the Tanner graph, as well as the output corrections of BP in the original and permuted graph.

\begin{figure}[h]
    \centering
    \includegraphics[width=\linewidth]{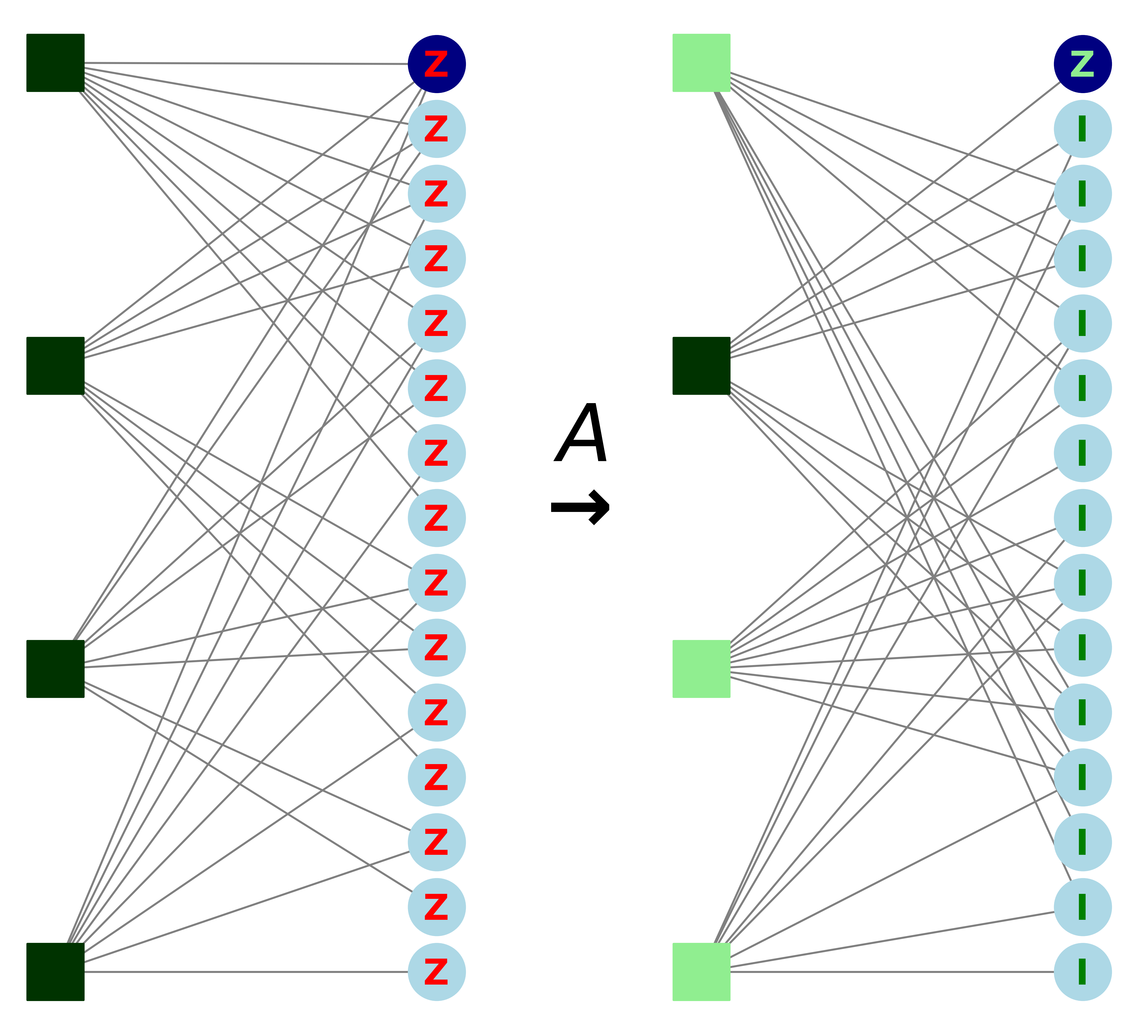}
    \caption{Left: Original Tanner graph of the $X$ checks of QRM$[[15,1,3]]$. An error on the $15$th qubit leads to a logical error that also does not return to the codespace. \\
    Right: The permuted Tanner graph. An error on the $15$th qubit is now corrected by BP.}
    \label{fig:bp_rm_15_aut}
\end{figure}
\end{example}

AutDEC is a heuristic algorithm and the performance varies depending on the automorphism group or  structure of the stabiliser code. 
A large automorphism group is generally a good indicator for potential performance gains. 
However, different decoders \enquote{absorb} different automorphisms \cite{geiselhart_polar_automorphism_2021, bioglio_group_2023, geiselhart_automorphism_2022}, or in other words, they provide equivalent corrections on different input syndrome data. 

\subsection{Description of AutDEC Decoder}

There are two components to the decoder:
\begin{enumerate}
    \item The offline setup where we prepare the decoder ensemble. 
    \item The online phase where we process the received syndrome data and output a correction.
\end{enumerate}

\subsubsection{Offline Setup of Ensemble Decoders}
We first outline the offline setup of the Automorphism Ensemble: 
\begin{enumerate}
    \item Find generators for the automorphism group of the code or of the parity check matrix. 
    For small codes we use the code automorphisms and for large codes, we use the graph automorphisms of the Tanner graph using Bliss \cite{junttila_engineering_2007, junttila_conflict_2011}. 
    We use a randomly selected set of automorphisms for ensemble decoding.
    For circuit-level noise, we use automorphisms of the detector error model check matrix (see \cite{derks_designing_2024}).

    \item Choose ensemble size. 
    In general a larger automorphism ensemble gives better performance and higher probability of convergence for each of the constituent BP decoders, but this may be limited by the classical hardware's distributed capabilities.  

    \item Find action of automorphism on checks of the code and syndrome vector. 
    For each automorphism $A$ in the set, we calculate the operator $U_A$ of Equation \ref{UA} which transforms both the parity check matrix and the syndrome vector using algorithm \ref{alg:stab_map}.

    \item Initialize a BP decoder for each automorphism in the ensemble based on the transformed parity check matrix. 
\end{enumerate}

\subsubsection{Online Ensemble Decoding}

\begin{figure}[H]
    \centering
    \includegraphics[width=1\linewidth]{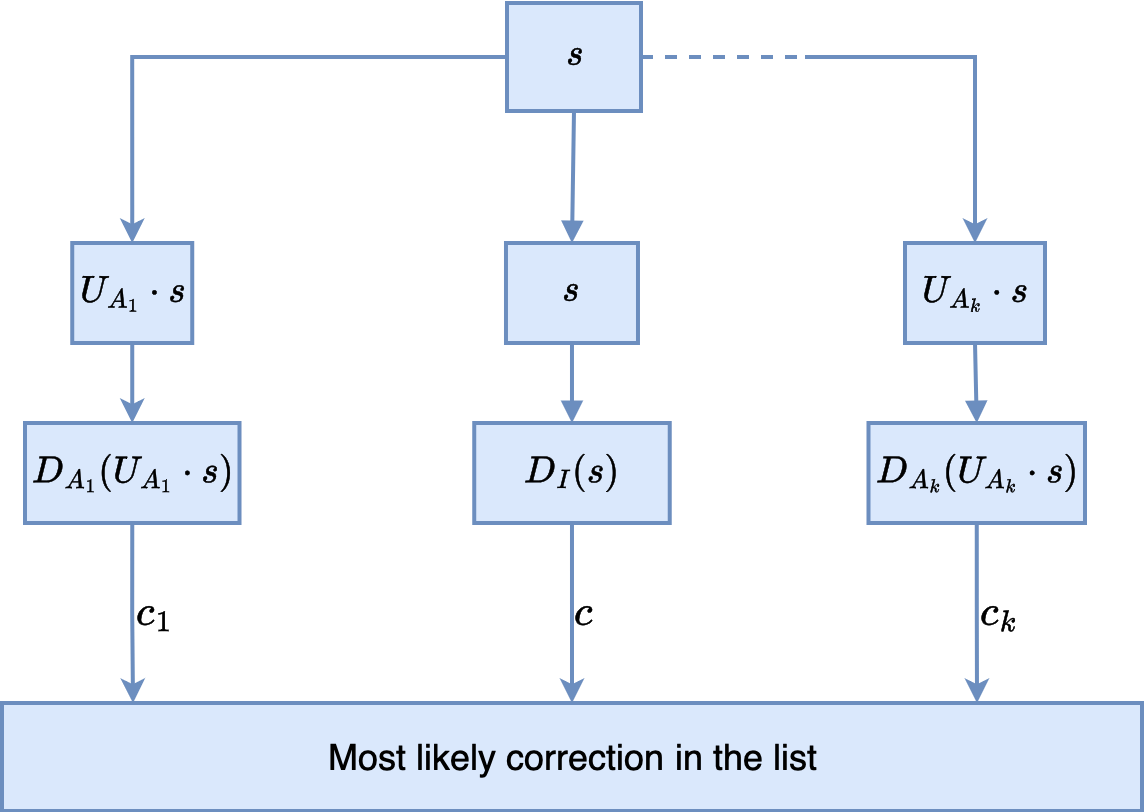}
    \caption{Overview of the AutDEC online decoding phase. Upon receiving the syndrome \textbf{s}, $k$ paths are created. In each path, the syndrome is left-multiplied by the $r\times r$ binary linear matrix $U_{A_i}$ corresponding to the action of the automorphism $A_i$ on the stabiliser generators. The syndromes are then sent to the $k$ constituent decoders $D_{A_i}$, each of which has been pre-compiled with the permuted check matrix $HA_i $. The corrections are gathered in a list, and the most likely correction in the list is then chosen based either on the minimum weight or the priors of the detector error model.}
    \label{fig:AED-overview}
\end{figure}

We now describe the operation of the AutDEC decoder for detection and correction of errors in the online phase.
The process is described graphically in Figure \ref{fig:AED-overview}.

\begin{enumerate}
    \item Measure the stabiliser generators and find the syndrome vector $s$.
    \item For each automorphism $A$, transform the syndrome vector using the operator of Equation \ref{UA} as follows $s_A = U_A\cdot s$

    \item Decode the transformed syndromes using the ensemble decoders.

    \item Form a list of candidate corrections based on the outcomes of the ensemble decoders which converge to a valid correction.

    \item Sort the candidate corrections based on a most likelihood metric which depends on the error model used. 
    For the code capacity setting (see \ref{sec:motivation}), we select the minimum weight Pauli operator in the correction list. 
    For circuit level noise (see \ref{sec:BB}), we sort the correction list based on the probability from the pre-computed priors given during the compilation of the detector error model. 
\end{enumerate}

\section{Results}\label{sec:results}

\subsection{[[15,1,3]] Quantum Reed Muller Code - Code Capacity}\label{sec:reed-muller}

In this section, we apply AutDEC to the $[[15,1,3]]$ Quantum Reed-Muller (QRM) code using a code capacity depolarising error model.
In the code capacity error model, we assume that errors are distributed on data qubits independently and identically with probability $p$ and that there are no correlated errors. 
We do not consider the effect of errors which occur during encoding, decoding, the application of gates or measurements.
Since the code is relatively small, we calculate the code automorphisms using the AutQEC package of \cite{Sayginel_autqec_Logical_Clifford_2024}.

\begin{figure}[H]
    \centering
    \includegraphics[width=\linewidth]{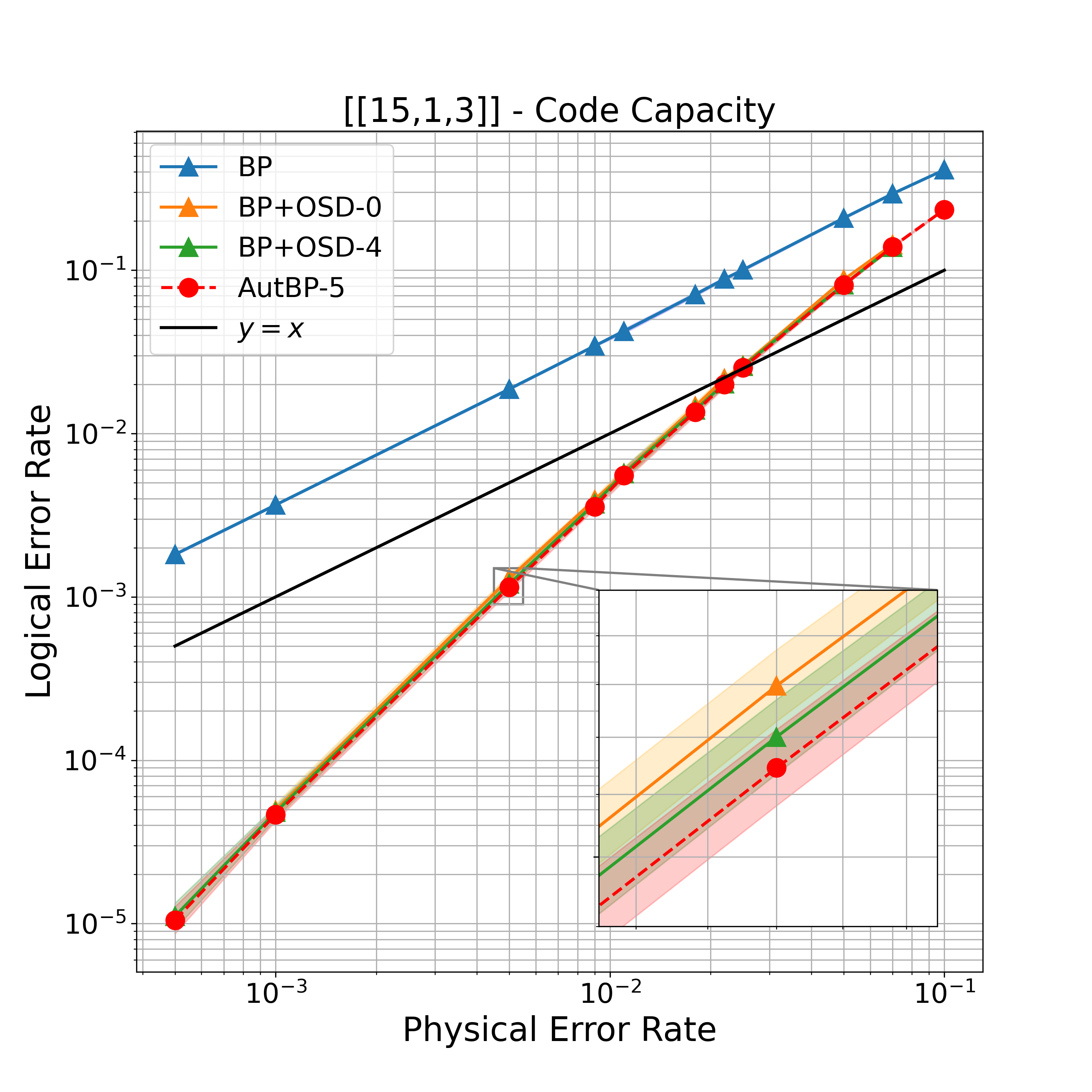}
    \caption{Logical error rate for the code capacity setting of the $[[15,1,3]]$ QRM code. We compare BP and BP+OSD-0 with AutBP-5. The black solid line is the $y=x$ break-even point. The shaded regions indicate the Wilson confidence interval of $95$\%.}
    \label{fig:rm15autbp}
\end{figure}

We use the LDPC package in \cite{Roffe_LDPC_Python_tools_2022} for the constituent BP and OSD decoders. 
We compare the performance of BP, automorphism ensemble BP and BP with OSD-0 and OSD-4 \cite{panteleev_degenerate_2021}. 
Each decoder uses a maximum of $15$ BP iterations.
We denote with AutBP-$N$ the ensemble decoder with $N$ BP decoders in the ensemble each using a different automorphism. 
We provide results for AutBP-$5$ and compare with BP, BP+OSD-0 and BP+OSD-4 in Figure \ref{fig:rm15autbp}.

We notice that BP does not achieve a threshold for any of the physical error rates, as its performance is limited by the large number of small cycles in the Tanner graph of the code. AutBP on the other hand with just a few automorphisms achieves a threshold and matches BP with OSD. 

A notebook implementing the method available at: \url{https://github.com/hsayginel/autdec/tree/main/code_capacity}.

\subsection{Bivariate Bicycle Codes - Circuit Level noise}\label{sec:BB}
Bivariate Bicycle Codes are a promising QLDPC code family recently introduced with high thresholds and encoding rates \cite{bravyi_high-threshold_2024}.
Here we apply the AutDEC decoder using a more realistic circuit-level noise model.
The circuit level noise model takes into account errors occurring during state initializations, application of gates, measurements and idling locations of the circuit.
We  assign probabilities to each of them occurring on each operation of the syndrome extraction circuit. 

The \textbf{detector check matrix} $H_\text{DEM}$ tracks the propagation of errors through the syndrome circuit.
It maps circuit fault locations to \textbf{detectors}, which are linear combinations of check outcomes.  

\begin{figure}[H]
    \centering
    \includegraphics[width=\linewidth]{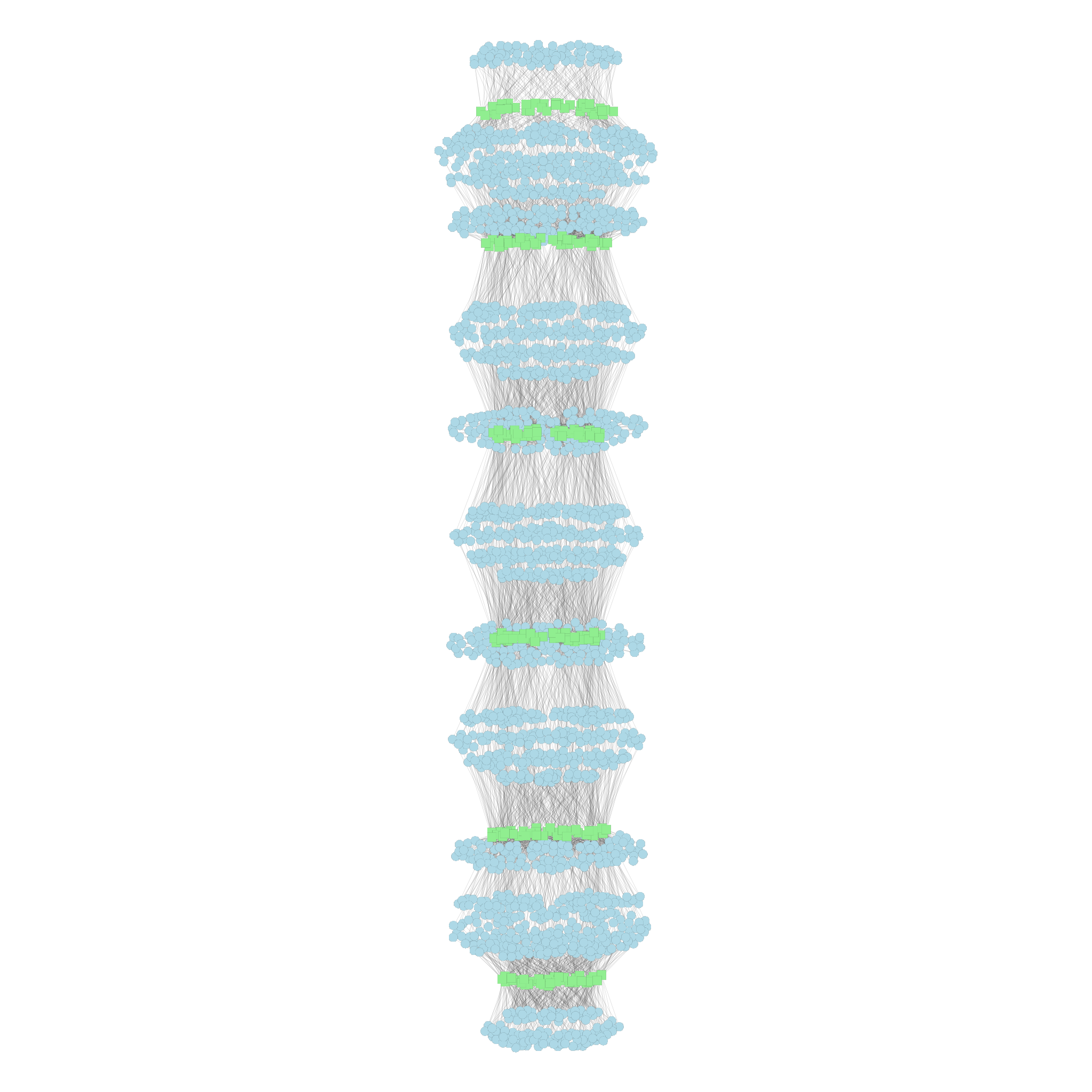}
    \caption{Visualisation of Tanner graph of $[[72,12,6]]$ $\ket{+_L}$ memory detector error model using Gephi \cite{bastian_gephi_2009}.}
    \label{fig:BB_DEM}
\end{figure}

The BP decoder can thus be used on the Tanner graph of the parity check matrix $H_\text{DEM}$ of the \enquote{circuit level code} instead. As these matrices are much larger than the parity check matrix of the original codes, finding  code automorphisms is  computationally expensive. 
In figure \ref{fig:BB_DEM}, we illustrate the $H_\text{DEM}$ for the $72$-qubit bivariate bicycle code which has $2,484$ vertices and $7,776$ edges. 
Finding the graph automorphisms of their Tanner graph, however, is feasible and we give processing times for the BB family in Table \ref{tab:automorphism_results}.
An in-depth overview of the detector error matrix construction is given in \cite{derks_designing_2024}.

We use the stabiliser measurement schedule described in \cite{bravyi_high-threshold_2024} implemented on stim \cite{gidney2021stim} in the GitHub repository of \cite{gong_toward_2024,gong_gongaaslidingwindowdecoder_2025}. This corresponds to a $Z$ memory experiment where the syndrome is extracted $d$ times for a $[[n,k,d]]$ code. For the simulation, the $d$ rounds are followed by a noiseless round of measurement of the data qubits. 

An interactive notebook is available at: \url{https://github.com/hsayginel/autdec/blob/main/bivariate_bicycle_codes/circuit_level_noise.ipynb}.

\begin{figure}[H]
    \centering
    \includegraphics[width=\linewidth]{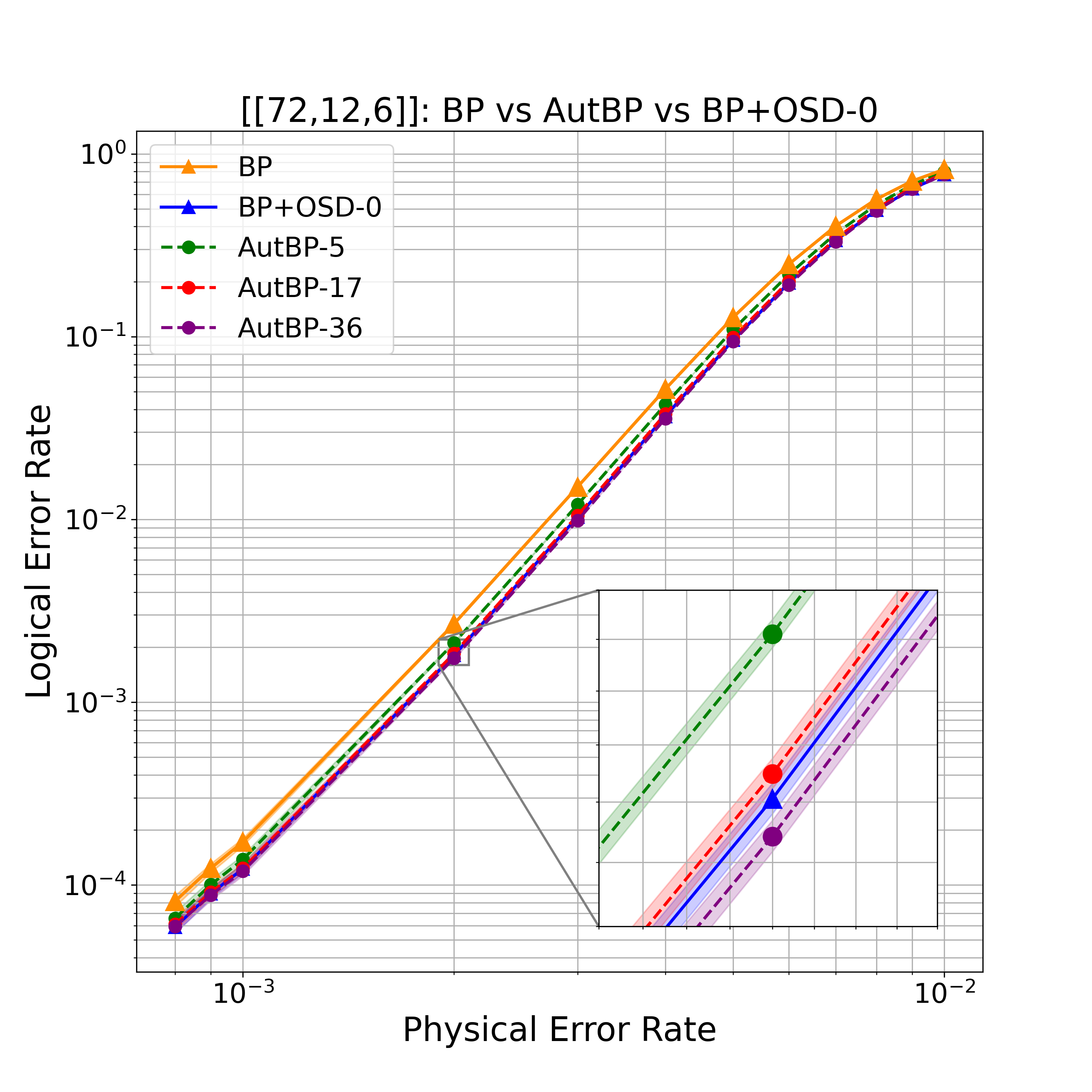}
    \caption{Logical error rate for different physical error rates using BP, AutBP-5, AutBP-17, AutBP-36 and BP+OSD-0 for circuit-level noise simulation of the BB [[72,12,6]] code. The BP decoders run a maximum of $1000$ iterations on a parallel schedule, with ``min-sum" scaling factor of 1.  The shaded regions indicate the Wilson confidence interval of 95\%.}
    \label{fig:bb72_autbp_vs_osd0}
\end{figure}

We first keep the base code fixed, and plot different ensemble sizes for the BP based automorphism ensemble decoder. We compare AutDec to applying a BP decoder on the original Tanner graph, as well as with using the quick ordered statistics decoding OSD-0 described in \cite{panteleev_degenerate_2021}. We plot the results in Figure \ref{fig:bb72_autbp_vs_osd0}.

We notice that increasing the ensemble size allows for the automorphism ensemble decoder to approach, match or marginally outperform OSD-0 using only BP constituent decoders.

\begin{figure*}[t]
    \centering
    \makebox[\textwidth]{ 
        \includegraphics[width=1\textwidth]{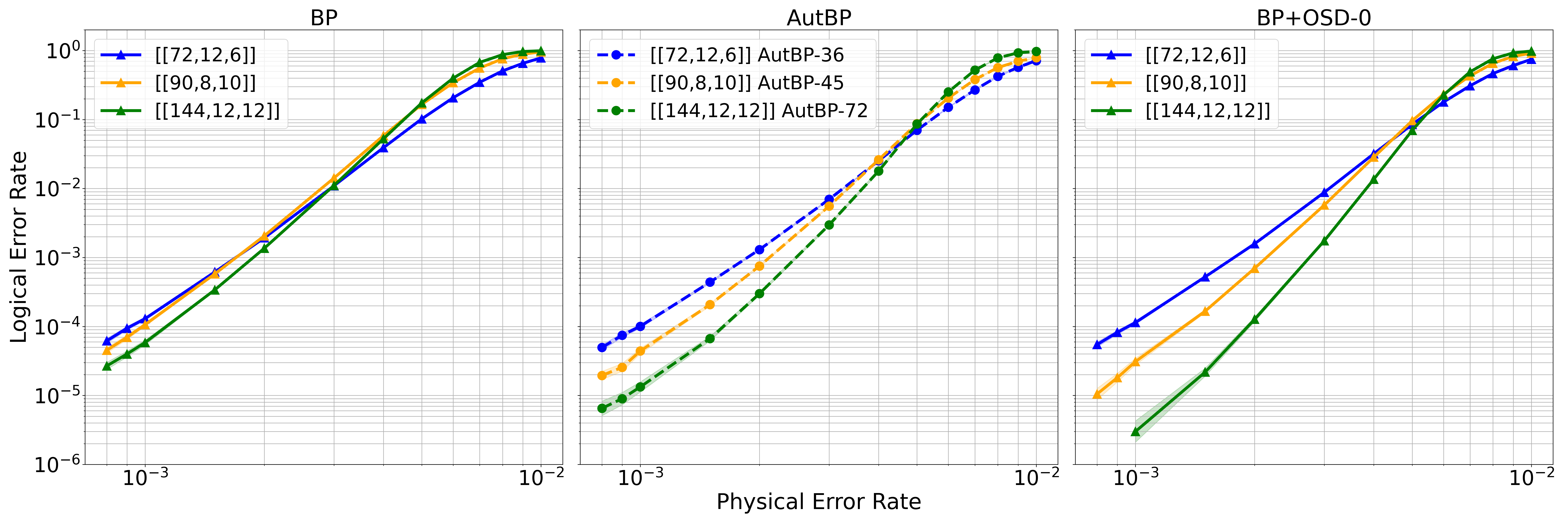} 
    }
    \caption{Logical error rate comparison between BP, AutBP and BP+OSD0 decoders for varying physical error rates. The BP decoders run a maximum of 10,000 iterations with parallel schedule and minimum sum updates. The minimum-sum scaling factor is 1 and in each case we have $d$ rounds of syndrome extraction in the Detector Error Model. The shaded regions indicate the Wilson confidence interval of 95\%.}
    \label{fig:bb_all}
\end{figure*}

Furthermore, in Figure \ref{fig:bb_all} we plot the logical error rate performance of the $[[72,12,6]], [[90,8,10]] $ and $[[144,12,12]]$ Bivariate Bicycle codes using only BP on the original Tanner graphs on the left panel, an automorphism ensemble based on BP using the full graph automorphism group of each code in the middle panel as well as using OSD-0 in the right panel. 

We notice that BP gets progressively worse when increasing the code size, as more and more cycles are introduced on the Tanner graph. The automorphism ensemble, however, bypasses that issue by using the Tanner graph automorphisms, approaching or matching the performance of OSD-0, without the cubic postprocessing overhead.

\begin{figure}[H]
    \centering
    \includegraphics[width=\linewidth]{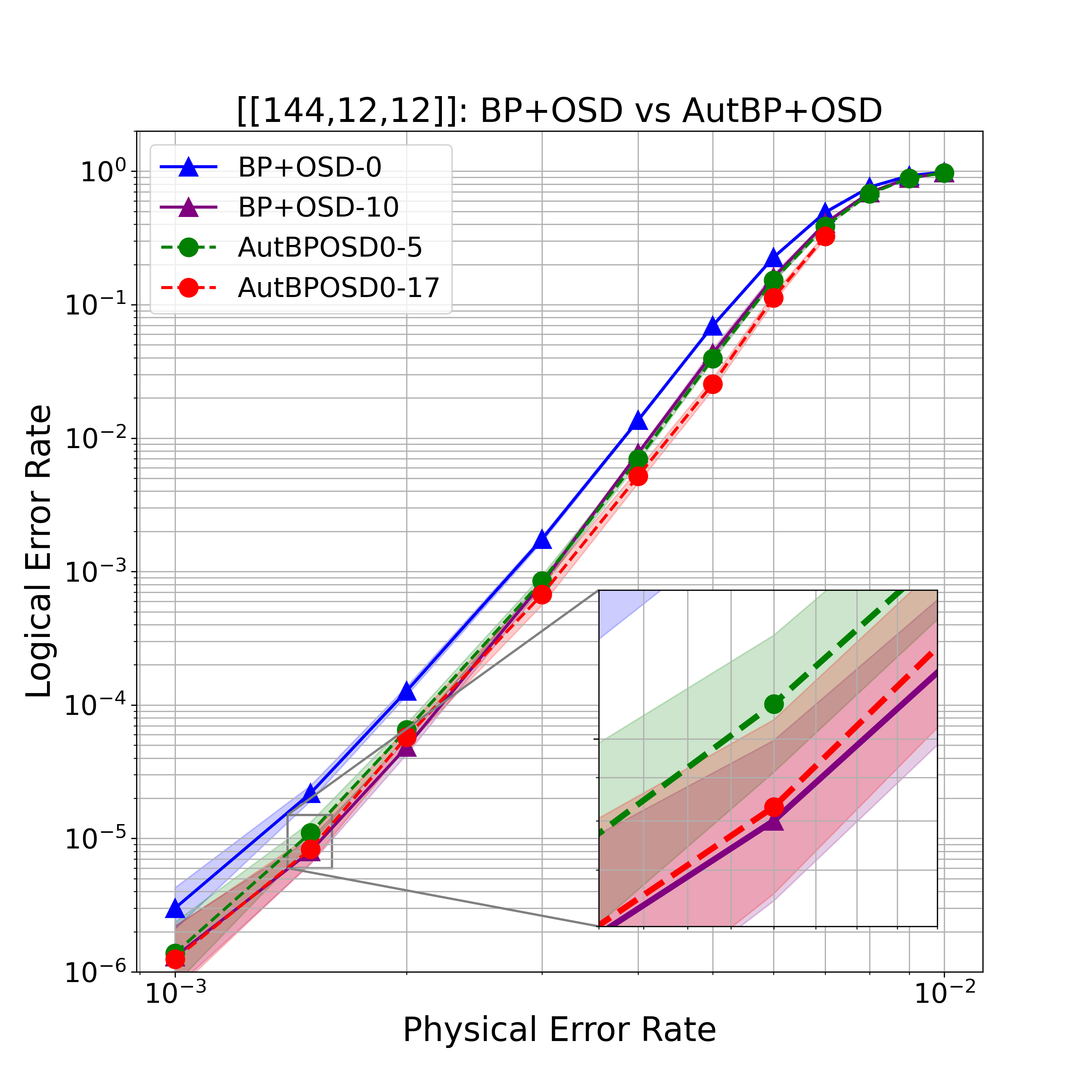}
    \caption{BB-[[144,12,12]]: Logical error rate comparison between BP+OSD-0, BP+OSD-10 and AutBPOSD0 with an automorphism ensemble size of 5 and 17. The BP decoders have a maximum of 10,000 iterations and a ``combination sweep" method is used for the OSD postprocessing. The shaded regions indicate the Wilson confidence interval of 95\%.}
    \label{fig:n144_cln_autbposd}
\end{figure}

Finally, in Figure \ref{fig:n144_cln_autbposd} , we plot the logical error rate of the $[[144,12,12]]$ Bivariate Bicycle code when using OSD-0 constituent decoders with different automorphism ensemble sizes. We compare it to the base BP with OSD-0, as well as to BP followed by OSD with combination sweep of order 10 postprocessing. We denote with AutBPOSD0-$N$ the ensemble decoder with $N$ BP+OSD0 decoders in the ensemble, each guided by a different automorphism. We compare BP+OSD-0, BP+OSD-10 with AutBPOSD0-5 and AutBPOSD0-17. 

We note that with fast OSD-0 constituent decoders, the automorphism ensemble yields similar or increased performance to higher OSD orders using only OSD0. 

The results of our simulations can be found at: \url{https://github.com/hsayginel/autdec/blob/main/bivariate_bicycle_codes/bb_cln_data.csv}.

\section{Conclusion and future work}
In this work, we introduced the AutDEC automorphism ensemble decoder. This approach leverages the inherent symmetries present in quantum error-correcting codes to enhance decoding performance.  
By employing multiple decoders in parallel, each guided by a distinct automorphism of the code, we improve the likelihood of successful BP decoding without the use of complex post-processing.

Our open source \href{https://github.com/hsayginel/autdec}{GitHub repository} \cite{Sayginel_AutDEC_2025} implementing AutDEC has been made available for use by the community.  
This implementation serves as a proof of concept, demonstrating the fundamental principles of the proposed method, as each automorphism-guided decoding path is executed sequentially, rather than concurrently. 
In future iterations, we would prioritize implementing full parallelization as well as exploration of larger, more diverse ensembles of decoders.

The potential incorporation of generalized automorphisms and endomorphisms, as recently investigated in \cite{mandelbaum_generalized_2023},\cite{mandelbaum_endomorphisms}, is another interesting direction. These more general symmetry mappings may enable the application of this method to codes with limited permutation automorphisms, although they come with an increased search space for relevant operations.

Finally, a deeper understanding of how code symmetries, as captured by automorphisms, influence decoding performance could potentially inform the design of new codes that are inherently more amenable to ensemble-based decoding techniques. Our work suggests that code automorphisms are tied both to decoding performance and low-overhead logical gate implementations \cite{sayginel_fault-tolerant_2024}, highlighting the fundamental relationship between code structure and error correction capabilities.

\begin{acknowledgments}
SK and DEB are supported by the Engineering and Physical Sciences Research Council [grant number EP/Y004620/1 and EP/T001062/1]. HS is supported by the Engineering and Physical Sciences Research Council [grant number EP/S021582/1]. HS also acknowledges support from the National Physical Laboratory. MW and DEB are supported by the Engineering and Physical Sciences Research Council [grant number EP/W032635/1 and EP/S005021/1]. The authors acknowledge the use of the UCL Myriad High Performance Computing Facility (Myriad@UCL), and associated support services, in the completion of this work. The authors would like to thank Oscar Higgott, Joschka Roffe and Tim Chan for helpful discussions. 
\end{acknowledgments}

\newpage

\bibliography{bibliography}

\clearpage
\onecolumngrid
\appendix

\section{Determining action of automorphism on parity checks}
We provide a more detailed description of our method for determining the action of a code automorphism  on the X-checks $H_X$ of a CSS code. 
Code automorphisms of CSS codes are qubit permutations $A$ which swap the columns of the check matrix.
They induce a linear map $U_A$ such that $U_A H_X = H_X A$.
Importantly, $U_A$ also transforms syndrome vectors $s$ of the check matrix to syndrome vectors $s_A = U_A\cdot s$ of the transformed check matrix.

\begin{algorithm}[H]
\caption{stabiliser Map Action Corresponding to Automorphism }
\label{alg:stab_map}
\KwIn{\begin{tabular}{@{}l@{}}
$H_X$: Original Parity check matrix of size $m \times n$ \\  $A$: a column permutation satisfying $\braket{H_A} = \braket{H_X}$.
\end{tabular}}
\KwOut{An invertible $m\times m$ binary matrix $U_A$ such that $H_XA = U_AH_X$.}
\BlankLine
Compute the reduced row echelon form (RREF) of $H_X$ modulo 2.\\
Store the pivots $P$, their indices $J$ and the row transformations matrix $R$ from  RREF.\\
Create a matrix $M$ of size $m \times m$. \\
Initialize all elements of $M$ to $0$.\\
\For{each row $i$ in $H_XA$}{
  \For{each index $j$ and corresponding pivot $p$ from the RREF}{
    \If{$HX_{new}[i,p]=1$}{
      Set $M[i, j] = 1$.
    }
  }
}
Multiply the $M$ matrix by the row transformation matrix (from RREF) modulo 2 to get the final stabiliser map matrix $U_A = M \times R$.\\
Return the resulting matrix $U_A$.
\end{algorithm}

Specifically for graph automorphisms, the matrix $U_A$ will be a row permutation matrix and can be directly extracted from the output of the graph automorphism. 

\subsection{Tanner Graph Automorphism Timings with (Py)Bliss}

In Table \ref{tab:automorphism_results} we provide an overview of the time required to find the automorphism group of the Tanner graphs associated with each of the detector error models of the bivariate bicycle codes. We use Bliss \cite{junttila_engineering_2007, junttila_conflict_2011} through the pyBliss \cite{kulkarni_pybliss_2025} wrapper. The reported results are from an Apple M3 8-core chip.

\begin{table}[htbp]
\centering
\label{tab:automorphism_results}
\begin{tabular}{|l|c|c|c|}
\hline
Code & DEM pcm shape & \begin{tabular}{@{}c@{}}Tanner graph automorphism \\group order \end{tabular} &  time (sec) \\
\hline
$[[72,12,6]]$ & $(252, 2232)$ & $36$ & $0.002$ \\
$[[90,8,10]]$ & $(495, 4590)$ & $45$ & $0.004$ \\
$[[108,8,10]]$ & $(594, 5508)$ & $54$ & $0.005$ \\
$[[144,12,12]]$ & $(936, 8784)$ & $72$ & $0.009$ \\
$[[288,12,18]]$ & $(2736, 26208)$ & $144$ & $0.024$ \\
$[[360,12,<=24]]$ & $(4500, 43560)$ & $180$ & $0.048$ \\
$[[756,16,<=34]]$ & $(13230, 129276)$ & $378$ & $0.139$ \\
\hline
\end{tabular}
\caption{Detector error model parity check matrices, Tanner graph automorphism group orders and time taken to find them using Bliss.}
\end{table}

\end{document}